# Depth-dependent hysteresis in adhesive elastic contacts at large surface roughness

Weilin Deng[1] and Haneesh Kesari[1,*]

[1]Brown University, School of Engineering, Providence, RI 02912, USA
*haneesh_kesari@brown.edu

**ABSTRACT**

Contact force–indentation depth measurements in contact experiments involving compliant materials, such as polymers and gels, show a hysteresis loop whose size depends on the maximum indentation depth. This depth-dependent hysteresis (DDH) is not explained by classical contact mechanics theories and was believed to be due to effects such as material viscoelasticity, plasticity, surface polymer interdigitation, and moisture, *etc*. It has been observed that the DDH energy loss initially increases and then decreases with roughness. A mechanics model based on the occurrence of adhesion and roughness related small-scale instabilities was presented by one of the authors for explaining DDH. However, that model only applies in the regime of infinitesimally small surface roughness, and consequently it does not capture the decrease in energy loss with surface roughness at the large roughness regime. We present a new mechanics model that applies in the regime of large surface roughness based on the Maugis–Dugdale theory of adhesive elastic contacts and Nayak's theory of rough surfaces. The model captures the trend of decreasing energy loss with increasing roughness. It also captures the experimentally observed dependencies of energy loss on the maximum indentation depth, and material and surface properties.

## Introduction

A clear understanding of adhesive contact mechanics is critical for spatially mapping out a material's mechanical properties using, *e.g.*, nanoindentation- and contact mode atomic force microscopy (AFM)-based techniques[1,2]. Typically, material properties are measured by fitting contact force vs. indentation depth (*P–h*) measurements to a contact mechanics theory. Some of the most popular theories for modeling adhesive elastic contact include the Johnson–Kendall–Roberts (JKR)[3], the Derjaguin–Muller–Toporov (DMT)[4], and the Maugis–Dugdale (MD)[5] theories. These classical contact theories predict that when the solids are in physical contact, the force is uniquely determined by the indentation depth and is independent of the history of the contact process [see Fig. 1 (a)]. However, in many experiments it is found that the contact forces depend on the contact process history. A typical contact experiment consists of one or more contact cycles, each of which consist of a loading and an unloading phase. In those phases the solids are, respectively, being moved towards and away from each other [Figs. 1 (a–b) and 2 (a)]. It is found that, at a given indentation depth, the contact force differs depending on whether the experiment is in a loading or an unloading phase [see Figs. 1 (b) and 2 (a)]. For example, Kesari *et al*.[6] reported AFM-based contact experiments between a glass bead and a poly(dimethylsiloxane) (PDMS) substrate, which shows that the contact forces differ between the loading and unloading phases [Fig. 2 (a)]. The force during the unloading phase was also observed to depend on the maximum indentation depth $|h_{\min}|$ [Fig. 2 (a)]. Kesari *et al*. termed this phenomenon depth-dependent hysteresis (DDH). The maximum indentation depth in a contact experiment is the indentation depth at the beginning of its unloading phase.

Depth-dependent hysteresis has also been observed in a number of other contact experiments, which span various length scales from $\mu$m to cm and involve different soft materials such as gelatin, PDMS, and poly(n-butyl acrylate) (PNBA)[7–9]. When the solids are in contact, the classical contact theories predict a single *P–h* curve [Fig. 1 (a)], whereas in the presence of DDH the experimental measurements display a different *P–h* branch for the loading and unloading phases [Figs. 1 (b) and 2 (a)], respectively. The estimates for the material properties are different depending on which branch is chosen to be fitted to a classical contact theory. For example, fitting the unloading and loading branches of the *P–h* curves shown in Fig. 2 (a) to the JKR theory yields values of 20 and 30 mJ/m$^2$, respectively, for the Dupré's work of adhesion *w*. Here $w = \gamma_1 + \gamma_2 - \gamma_{12}$, where $\gamma_1$ and $\gamma_2$ are the surface energies of the two solids, respectively, and $\gamma_{12}$ is the interfacial energy[10]. In some experiments, the ambiguity in the estimated values for *w* can be quite dramatic. For example, the *P–h* measurements reported by Guduru *et al*.[11] for contact between a polycarbonate punch and a gelatin slab display significant DDH, with the measurements falling into distinct loading and unloading branches. Fitting the loading branch of those measurements to the JKR theory yields a value of 8 mJ/m$^2$ for *w*, whereas fitting the unloading branch yields a value of 220 mJ/m$^2$.

Depth-dependent hysteresis has been attributed to various mechanisms, such as the meniscus effect of ambient moisture[9], the entanglement and interdigitation of tethered chains[12], the formation of hydrogen bonds[13], and the inelastic behaviors of materials (viscoelasticity[14] and plasticity[15]). However, Kesari *et al.*[6] showed that DDH persists even when the aforementioned mechanisms can be reasonably excluded. Motivated by the observation of overlapping hysteresis loops during consecutive load-unload cycles both in air and underwater, they hypothesized that DDH was due to the occurrence of a series of small-scale surface, mechanical instabilities that are created due to surface roughness, adhesion, and the large compliance of the soft materials involved[16]. Our recent static molecular simulations showed that this mechanism can operate in adhesive elastic contacts[17]. The surface instabilities through which small-scale roughness gives rise to DDH in the work of Kesari *et al.*[6] and Kesari and Lew[16] are the same as those through which surface undulations cause adhesive toughening in the work of Li and Kim[18] and Guduru[19].

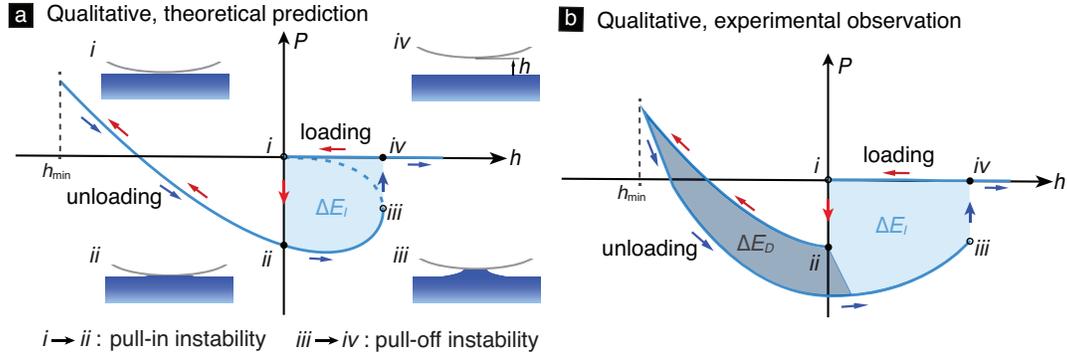

**Figure 1.** (a) The schematic of the *P–h* curve as per the JKR theory. The "pull-in" ($i \to ii$) and "pull-off" ($iii \to iv$) instabilities are marked along with the corresponding contact configurations. Closed and open symbols (circles) mark stable and unstable states on the *P-h* curve, respectively. A contact cycle includes the loading (red arrows) and unloading (blue arrows) phases. The size of the hysteresis loop formed in a contact cycle due to the instabilities (*i.e.*, the shaded area $\Delta E_I$) denotes the hysteretic energy loss, which is depth independent. (b) The schematic of the *P-h* curve observed in some experiments [*e.g.*, see Fig. 2 (a)] which shows that the contact forces differ between the loading and unloading phases. The total hysteretic energy loss includes depth-independent part $\Delta E_I$ and depth-dependent part $\Delta E_D$.

The area enclosed by the *P-h* curves in a contact cycle, $\Delta E$, is a measure of the energy lost during that cycle. It was found experimentally that $\Delta E$ initially increases and then later decreases with the surface roughness[6,20], *e.g.*, see Fig. 2 (b). Kesari *et al.*[6,16] presented a model for DDH that captures many of the salient features of DDH, including the initial increase of $\Delta E$ with the root mean square (RMS) roughness $\sigma$. However, that model does not capture the later decreases of $\Delta E$ with $\sigma$. Kesari *et al.* and we believe that this fact is due to the model's assumption that the contact region is simply connected [top inset, Fig. 2 (b)]. The contact region between two flat, perfectly smooth surfaces would be simply connected. It is likely to remain so even if infinitesimally small undulations were superimposed onto the flat surfaces. This would be especially true if the solids were composed of compliant materials, such as hydrogels or nonmineralized, biological tissues. However, irrespective of the compliance of the materials, as the height of the undulations is increased and the surface becomes rougher, the contact region will eventually become multiply connected [bottom inset in Fig. 2 (b)].

In this work, we focus on the regime of large surface roughness where the contact region is multiply connected, and present a new model that captures the trend of $\Delta E$ decreasing with $\sigma$. This model is based on the MD theory of adhesive elastic contacts and the Nayak's theory of rough surfaces[21]. The mechanism of energy loss in this model is similar to the one in the model presented by Kesari and Lew *et al.*[16], in which the energy loss arises as a consequence of small-scale surface mechanical instabilities. The primary difference between the model presented in[16] and the new model herein is that the contact region in the former is simply connected whereas in the latter it is multiply connected.

Our new model involves adhesive elastic contact between a smooth, rigid paraboloid (tip) and a rough, semi-infinite, deformable solid (substrate) [see Fig. 6 (a)]. The substrate's surface facing the tip is nominally flat but contains a random distribution of asperities. There are two major types of models used for studying contact between rough surfaces. The first type is based on the *non-interacting asperity contact model* pioneered by Greenwood and Williamson[22], which is widely used for studying the effect of roughness on adhesion[23,24], particle adhesion[25], elasto-plastic contact[26], and friction[27]. The second type is related to the *self-affine fractal contact model* put forward by Persson[28]. Ours is a non-interacting asperity type contact model, in which we assume that each substrate asperity interacts with the tip as though it were the only one interacting with it.

The energy loss $\Delta E$ was found experimentally to scale affinely with $|h_{\min}|$, with its minimum value corresponding to the case $|h_{\min}| = 0$. Furthermore, it was found that $\Delta E$ can be partitioned into two parts: a fixed part $\Delta E_I$ that only depends on the geometry and mechanical properties of the contacting solids and not on $|h_{\min}|$ and a variable part $\Delta E_D$ that in addition to the solids' geometric and mechanical properties also depends on $|h_{\min}|$ [see Fig. 1 (b)]. The depth-independent part $\Delta E_I$ is a consequence of two surface mechanical instabilities that occur at the large-scale. These large-scale instabilities correspond to the initial sudden drop in the contact force [*i.e.*, the transition from state (*i*) to (*ii*) in Fig. 1 (b)] and the final abrupt increase in contact force [*i.e.*, the transition from state (*iii*) to (*iv*) in Fig. 1 (b)]. These instabilities are generally referred to as "pull-in" and "pull-off" instabilities. It was observed in the experiments[6] that each of these large-scale instabilities always occurred once and only once in a contact cycle. Therefore, $\Delta E_I$ is the fixed, minimum amount of energy that gets dissipated in every contact cycle. Consequently, $\Delta E_I$ can be computed as the total energy dissipated in the contact cycle with $|h_{\min}| = 0$.

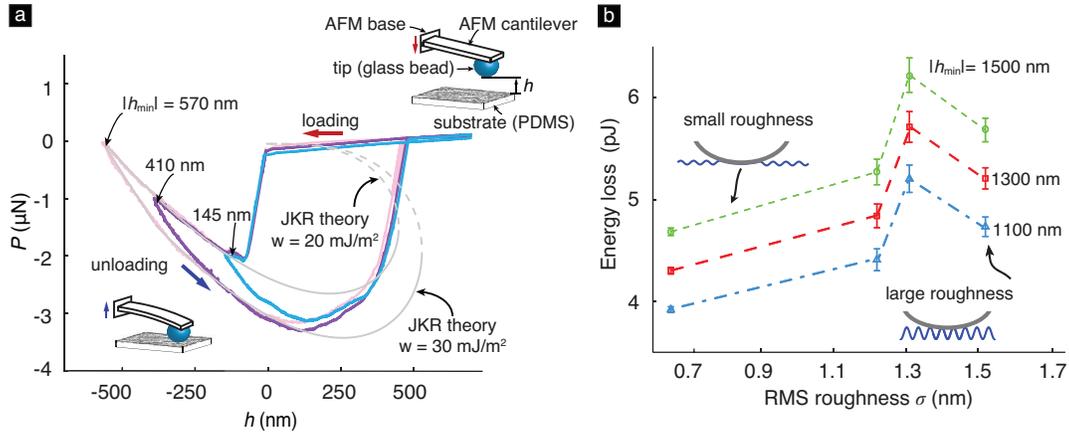

**Figure 2.** (a) Representative *P–h* curves measured in AFM contact experiments between a glass bead and a PDMS substrate[6]. The glass bead was of diameter $\approx 50~\mu$m. The PDMS sample was cast on a silicon wafer having a RMS roughness $\approx 1.3$ nm. As can be noted, the measured *P–h* curves for the loading and unloading phases of the experiment are different. The size of the hysteresis loop increases with the maximum indentation depth, $|h_{\min}|$. The gray dashed curves are the fit of the loading and unloading branches of the measured *P–h* data to the JKR theory. (b) A plot showing the variation of total energy loss as function of the RMS roughness in the experiments. The RMS roughness refers to the surface roughness of the silicon wafer on which the PDMS substrates were cast. The indenting rate in the experiments corresponding to all data points shown in the plot was 1000 nm/s. See Ref.[6] for experimental details.

After the occurrence of the large-scale "pull-in" instability, as the solids are moved towards one another, more and more surface asperities will come into contact. We assume that those surface asperities will come into contact through small-scale surface mechanical instabilities, as done in Ref[6, 16]. Consequently, we refer to the asperities that come into contact during the loading phase after the occurrence of the large-scale "pull-in" instability as the *depth-dependent asperities*. Classical contact theories, which ignore roughness, predict that the contact radius prior to the occurrence of the large-scale "pull-off" instability is smaller than that after the occurrence of the large-scale "pull-in" instability. We assume that that prediction holds true even in the presence of roughness and therefore that during the unloading phase there will be a point when the contact region has receded back—in a nominal (large-scale) sense—to the one formed just after the occurrence of the large-scale "pull-in" instability. This implies that all the depth-dependent asperities would go out of contact before the occurrence of the large-scale "pull-off" instability. We assume that the detachment of the depth-dependent asperities takes place through the occurrence of small-scale instabilities, too. Thus, the energy loss $\Delta E_D$ consists of the energy lost during the instabilities through which the depth-dependent asperities come into and go out of contact. Since the larger the $|h_{\min}|$ the larger will be the number of depth-dependent asperities, the loss $\Delta E_D$ increases with $|h_{\min}|$.

This paper is organized as follows: First, we evaluate the energy loss corresponding to the pair of small-scale "pull-in" and "pull-off" instabilities by using the MD theory; Second, based on the Nayak's theory of rough surfaces, we estimate the number of depth-dependent asperities and the depth-dependent energy loss $\Delta E_D$ during a contact cycle; Furthermore, we discuss the comparisons of the theoretical prediction of $\Delta E_D$ based on our model with the experimental measurements; Finally, we conclude by discussing the limitations of our model.

## Theory

### Energy loss per asperity using the Maugis-Dugdale theory

The MD theory describes the axi-symmetric contact between two isotropic, homogeneous, linear elastic solids of Young's moduli and Poisson's ratios $E_i$ and $\nu_i$ ($i = 1, 2$), respectively [see Fig. 3 (a)]. The adhesive interactions are introduced using the

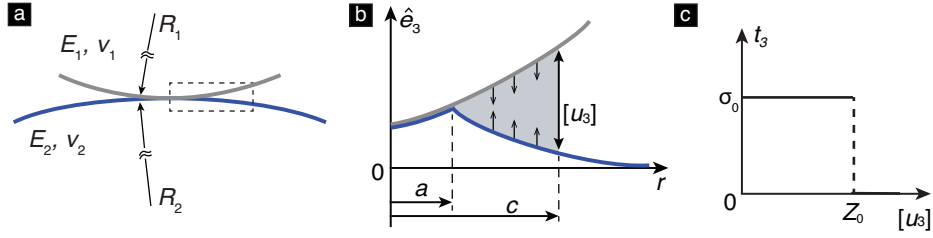

**Figure 3.** The MD model of adhesive elastic contact. (a) Geometry of the contacting solids. (b) The Dugdale cohesive zone, which is assumed to be present at the contact periphery (marked by the dashed box in (a)) as per the MD theory. The vector $\hat{\mathbf{e}}_3$ belongs to the set of Cartesian unit basis vectors, $\{\hat{e}_i\}_{i=1,2,3}$, which is defined in Fig. 6 (a). The symbol $r$ denotes the radial coordinate in the plane spanned by $\hat{\mathbf{e}}_1, \hat{\mathbf{e}}_2$. The datum of $r$ lies on the axial symmetry axis of the contacting solids. (c) A schematic diagram showing the traction distribution $t_3$ as a function of the separation $[u_3]$. The traction $t_3 = \hat{\mathbf{e}}_3 \cdot (\sigma \hat{\mathbf{e}}_3)$, where $\sigma$ is the Cauchy stress tensor. The parameters $Z_0$ and $\sigma_0$ are defined in the text.

*Dugdale cohesive zone* model[29]. As per this model, a surface material point experiences a traction only if its distance from the other solid in the direction normal to the surface is less than $Z_0$. Thus $Z_0$ denotes the range of the inter-body adhesive forces, which are thought to arise from van der Waals type interactions between the surfaces. When the normal distance of the material point is less than $Z_0$ but non-zero then the traction experienced by it is purely tensile and of a fixed magnitude of $\sigma_0$ [see Fig. 3 (b–c)].

The contact process is governed by the two dimensionless parameters

$$\ell_0 = \frac{\pi w}{E^* R} \tag{1}$$

and

$$\chi = \frac{w}{E^* Z_0}, \tag{2}$$

where $1/R = 1/R_1 + 1/R_2$ is the sum of the mean curvatures of the contacting solids at their respective points of contact, and $1/E^* = (1-\nu_1^2)/E_1 + (1-\nu_2^2)/E_2$. In terms of these non-dimensional parameters, the magnitude of the contact force $P$ and the indentation depth $h$ at equilibrium are related as

$$\begin{aligned}
\bar{P} &= \begin{cases} \frac{2}{3}\bar{a}^3 - \chi \bar{a}^2 \left(\sqrt{m^2-1} + m^2 \tan^{-1}\sqrt{m^2-1}\right), & \bar{a} > 0, \\ -\frac{\pi}{2}\chi \bar{c}^2, & \bar{a} = 0, \end{cases} \\
\bar{h} &= \begin{cases} -\bar{a}^2 + 2\chi \bar{a}\sqrt{m^2-1}, & \bar{a} > 0, \\ 2\chi \bar{c} + \bar{h}_g, & \bar{a} = 0, \end{cases} \\
\ell_0 &= \begin{cases} \bar{a}^2 \left[\sqrt{m^2-1} + (m^2-2)\tan^{-1}\sqrt{m^2-1}\right] \\ \quad + 4\bar{a}\chi^2 \left[\sqrt{m^2-1}\tan^{-1}\sqrt{m^2-1} - m + 1\right], & \bar{a} > 0, \\ \frac{\pi}{2}\chi \bar{c}^2 + 2(\pi-2)\chi^2 \bar{c} + \pi \chi \bar{h}_g, & \bar{a} = 0, \end{cases}
\end{aligned} \tag{3}$$

where $\bar{P} = P/(2E^* R^2)$, $\bar{h} = h/R$, $\bar{h}_g = h_g/R$, $\bar{a} = a/R$, $\bar{c} = c/R$ and $m = c/a$. The parameter $c$ is defined such that all surface points whose radial coordinate in the undeformed configuration, $r$, is less than or equal to $c$ experience a non-zero traction force [see Fig. 3 (b)]. The coordinate system corresponding to $r$ is defined in Fig. 3 and its caption. The parameter $a$ is defined such that there is no separation, $[u_3]$, between the solids' surfaces in the region $r \le a$. The separation $[u_3]$ is defined in Fig. 3 (b) and is usually referred to as the *crack opening displacement*[10]. The parameter $h_g$ is the separation between the solids' surface points at $r = 0$ when $a = 0$. Due to the finite range of the inter-body surface adhesive interactions, the surface tractions in the MD theory do not vanish when $\bar{a} \to 0$, which is the case in the JKR and Hertz theories. For this reason we refer to $c$ as the contact radius. The cases $\bar{a} > 0$ and $\bar{a} = 0$ in eq. (3) were, respectively, derived by Maugis[5] and Kim *et al.*[30]

Figure 4 shows the representative equilibrium $P$–$h$ curves for different combinations of parameters $\chi$ and $\ell_0$ according to eq. (3). When $\ell_0 > 0$, the MD theory asymptotes to the JKR and DMT theories, respectively, as $\chi \to \infty$ and 0 [Fig. 4 (a)]. The JKR theory applies to compliant materials having a large work of adhesion, while the DMT theory applies to stiff materials having a small work of adhesion. When $\chi$ is any finite, fixed value, then as $\ell_0 \to 0$ the MD theory asymptotes to the Hertz theory [Fig. 4 (b)].

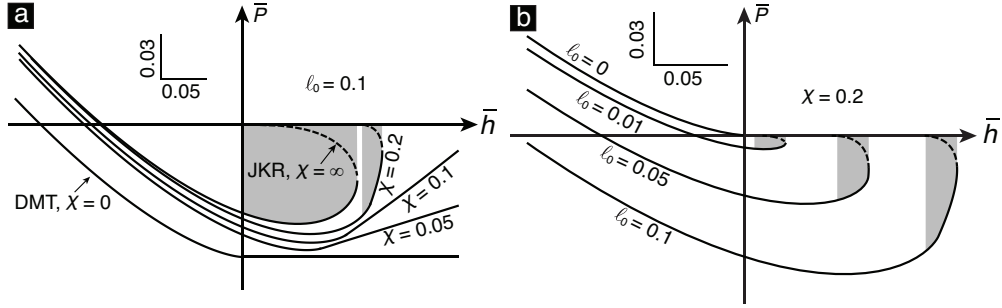

**Figure 4.** (a) The equilibrium *P-h* curves predicted by the MD theory for different $\chi$ values, with $\ell_0$ being held fixed at 0.1. The JKR and DMT limits are achieved when $\chi \to \infty$ and $\chi \to 0$, respectively. (b) The curves for different $\ell_0$ values, with $\chi$ being held fixed at 0.2. The Hertz limit is achieved when $\ell_0 \to 0$. In both plots, the solid and dashed segments denote stable and unstable equilibrium states, respectively. The shaded area indicates the energy loss of the hysteresis loop.

When $\ell_0 > 0$, the MD theory predicts that the solids will come into and go out of contact through the well-known mechanical instabilities termed the "pull-in" and "pull-off" instabilities during a contact cycle. The schematic of a typical equilibrium *P–h* curve predicted by the MD theory is shown in Fig. 1 (a). In that schematic, the "pull-in" and "pull-off" instabilities, respectively, correspond to the initial sudden drop in the contact force [state (*i*) to (*ii*)] and the final sudden increase in the contact force [state (*iii*) to (*iv*)]. In a displacement controlled experiment, the measured *P–h* curve will be the envelope of the equilibrium *P–h* curve. The energy lost during a contact cycle, $\Delta E_{\text{md}}$, due to the "pull-in" and "pull-off" instabilities, is equal to the area enclosed by the *P–h* curves measured during that cycle. It is denoted as the shaded area in Fig. 4.

The energy loss can be computed from the $\bar{P}$–$\bar{h}$ curves, which are defined by eq. (3), as

$$\Delta E_{\text{md}} = (2E^* R^3) \Delta \bar{E}_{\text{md}}, \tag{4a}$$

where

$$\Delta \bar{E}_{\text{md}} = \int_{r_i}^{r_o} \bar{P}(r) \frac{\partial \bar{h}(r)}{\partial r} dr, \tag{4b}$$

The limits of integration $r_i$ and $r_o$ in eq. (4b) are the contact radii at the instances just after and prior to the occurrence of the "pull-in" and "pull-off" instabilities, respectively. We refer to $\Delta \bar{E}_{\text{md}}$ as the normalized energy loss. Since the $\bar{P}$–$\bar{h}$ curves are completely defined by $\chi$ and $\ell_0$, the energy loss $\Delta \bar{E}_{\text{md}}$ only depends on these two parameters, too. We could not find a closed form expression for $\Delta \bar{E}_{\text{md}}$, by evaluating the integral in eq. (4b) analytically for arbitrary values of $\chi$ and $\ell_0$. However, we were able to obtain closed form expressions in three special cases. When $\chi \to \infty$, with $\ell_0$ held fixed, we find that $\Delta \bar{E}_{\text{md}} \sim 0.5262 \ell_0^{5/3}$. On the other hand, when $\chi \to 0$ with $\ell_0$ held fixed we obtain that

$$\Delta \bar{E}_{\text{md}} \sim 5.8483 \chi^5, \tag{5}$$

as shown in Fig. 5 (a). Finally, when $\ell_0 \to 0$ with $\chi$ held fixed, the energy loss $\Delta \bar{E}_{\text{md}} \to 0$.

We numerically compute $\Delta \bar{E}_{\text{md}}$ for a wide range of $\chi$ and $\ell_0$ values (see Fig. 5). As can be seen, $\Delta \bar{E}_{\text{md}}$ increases with both $\chi$ and $\ell_0$. By analyzing the numerical data shown in Fig. 5, we find that the dependence of $\Delta \bar{E}_{\text{md}}$ on $\chi$ and $\ell_0$ can be well approximated by the values of the empirical function

$$\Delta \tilde{E}_{\text{md}}(\chi, \ell_0) = \frac{c_1 \chi^5}{[c_2 (\chi^3 / \ell_0) + 1]^{5/3}}, \tag{6}$$

where $c_1 = 5.8483$ and $c_2 = 4.2415$. A comparison of the approximate values of $\Delta \bar{E}_{\text{md}}$ given by eq. (6) with its exact values computed numerically is shown in Fig. 5 (b). A notable aspect of the empirical function $\Delta \tilde{E}_{\text{md}}$ is that it gives the exact values of $\Delta \bar{E}_{\text{md}}$ in the limit $\chi \to 0$ and $\chi \to \infty$, while holding $\ell_0$ fixed, and also in the limit $\ell_0 \to 0$, while holding $\chi$ fixed. The differences between $\Delta \bar{E}_{\text{md}}$ and $\Delta \tilde{E}_{\text{md}}$ are more noticeable at intermediate values of $\chi$. However, we found those differences to be less than 15% for the data shown in Fig. 5. Therefore, we will approximate $\Delta \bar{E}_{\text{md}}$ with $\Delta \tilde{E}_{\text{md}}$ in our remaining analysis.

## Depth-dependent energy loss due to the asperity level instabilities

In this section we present a rough surface contact model, and use that model to compute the depth-dependent part of the energy loss, $\Delta E_D$, as the product of the total number of depth-dependent asperities and the mean energy loss per asperity.

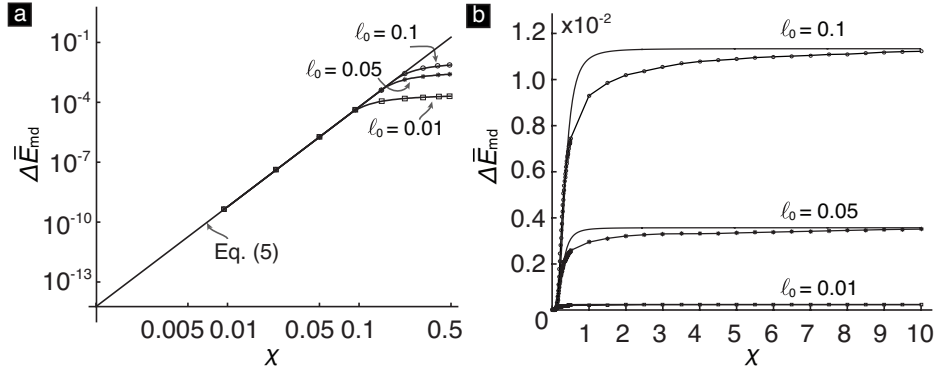

**Figure 5.** (a) The plot of the eq. (5) and the numerically computed $\Delta \bar{E}_{\text{md}}$ for different $\ell_0$ when $\chi$ is very small. (b) The comparison of exact (line with symbols) with approximate (solid line) values of $\Delta \bar{E}_{\text{md}}$ for different $\chi$ and $\ell_0$. The exact values are computed numerically using eqs. (3)–(4b). The approximate values are computed from eq. (6).

In our rough surface contact model, the tip is a paraboloid with the radial profile $\tilde{u}_3 = h + r^2/2R_t$, $r \in [0, R_t]$. We describe the geometry of our model using the Cartesian coordinates $x_i$, $i = 1, 2, 3$, whose corresponding basis vectors, $\hat{\mathbf{e}}_i$, are shown in Fig. 6 (a). We describe the substrate's surface topography using the function $z : \mathbb{R}^2 \to \mathbb{R}$, which gives the height ($x_3$ coordinate) of the substrate's surface points as a function of their $x_1$, $x_2$ coordinates. The datum of the $\hat{\mathbf{e}}_3$ direction is chosen such that $\int_{\mathbb{R}^2} z(x_1, x_2) \, dx_1 dx_2 = 0$. That is, the set of points $x_3 = 0$ form the mean plane of the substrate's rough surface [see Fig. 6 (c)]. The datums of the $\hat{\mathbf{e}}_1$, $\hat{\mathbf{e}}_2$ directions are chosen such that the coordinate system's origin is the point where the tip's rotational symmetry axis intersects the mean plane.

Consider a region in the mean plane having an area of unit magnitude. We say that this unit region contains an asperity whose apex has the co-ordinates $(x_1, x_2, z(x_1, x_2))$, if it contains the point $(x_1, x_2, 0)$. The unit region will, in general, contain a large number of asperities. A number of surface topography measurements have shown that the variation of a rough surface's geometric features can be well described using stochastic models[31,32]. Motivated by those results, we model the variation of the different geometric characteristics of the asperities belonging to the unit region using the probability density functions (PDFs) given by Nayak[21]. In our current model, we assume that, in a statistical sense, the substrate's surface roughness is homogeneous and isotropic. That is, the PDFs characterizing the different geometric features of the asperities do not depend on the location or the orientation of the unit region. For this special case, Nayak[21] gives the joint PDF of the heights and curvatures of the asperities belonging to the unit region to be

$$p(\xi, t) = \frac{\sqrt{3C_1}}{2\pi} e^{-C_1 \xi^2} \left( t^2 - 2 + 2 e^{-\frac{t^2}{2}} \right) e^{-\frac{C_1 t^2 + C_2 t \xi}{2}}, \tag{7}$$

where $\xi \in (-\infty, \infty)$ is the asperity height normalized by the surface's RMS roughness $\sigma$ [see Fig. 6 (c)], $t = -\sqrt{3/m_4} k_m$, and the asperity curvature $k_m \in (0, \infty)$ is the surface's mean curvature at the apex of an asperity [see Figs. 6 (c–d)]. The constants $C_1$, $C_2$ in eq. (7) are defined as $C_1 := \alpha/(2\alpha - 3)$ and $C_2 := C_1 \sqrt{12/\alpha}$, where $\alpha$ is an important parameter called Nayak's parameter. It is defined as $\alpha := m_0 m_4 / m_2^2$, where $m_0$, $m_2$, and $m_4$ are the surface's spectral moments. These moments can be computed from the equation

$$m_n = \frac{2\sqrt{\pi}\Gamma((1+n)/2)}{\Gamma(1+n/2)} \int_0^\infty q^{n+1} C^{\text{iso}}(q) \, dq \tag{8}$$

by setting $n = 0, 2$ and $4$, respectively, where the function $C^{\text{iso}} : \mathbb{R} \to \mathbb{R}$ is the isotropic surface's power spectral density (PSD). It is determined by the surface's topography $z$. See Supplementary Material for its complete definition.

We assume in our model that the contact between the tip and the substrate takes place only at the asperities. Consequently, in our model the real contact region is smaller than the nominal contact region. We define the nominal contact region to be a circular region in the mean plane that contains all the contacting asperities [Fig. 6 (b)]. The nominal contact region is also referred to as the apparent contact region in the literature. Since, at the large-scale it is the region over which the solids appear to be in contact. The nominal contact region grows and recedes during the loading and the unloading phases of the contact cycle, respectively. The evolution of the real contact region is much more complicated. The definition of the nominal contact region, by itself, does not imply that all asperities contained within it are in contact with the tip. Indeed, it is possible that many asperities never make contact despite belonging to the nominal contact region during some instance of the contact cycle.

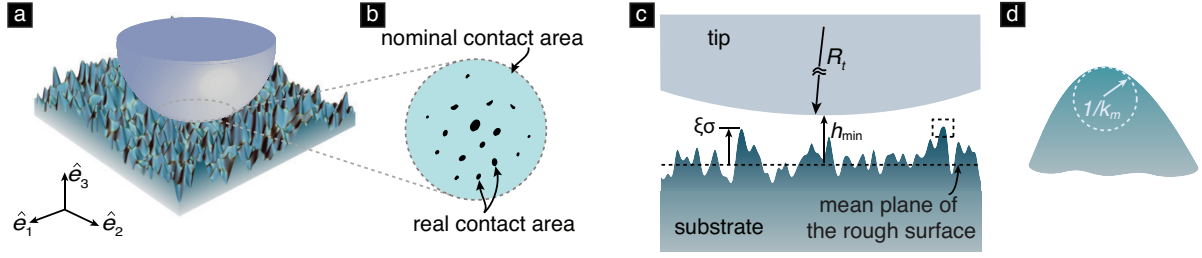

**Figure 6.** (a) The schematic of contact between a smooth rigid tip and a rough elastic substrate. (b) The schematic of nominal and real contact areas. (c) The section-view of the rough contact model shown in (a). (d) An asperity with radius of curvature $1/k_m$ at its apex as indicated by the dashed box in (c).

However, as part of our model, we assume that all asperities within a nominal contact region make and break contacts with the tip as that region forms and unforms. As a consequence of this assumption, the total number of depth-dependent asperities can be computed as the product of the asperity density and the area of the nominal contact region that forms after the occurrence of the large-scale "pull-in" instability during the reminder of the contact cycle's loading phase. The asperity density is the total number of asperities contained in a nominal contact region of unit area. Nayak[21] gives the total number of asperities contained in a region of the mean plane of unit area to be

$$\eta = \frac{m_4}{6\pi\sqrt{3}m_2}. \tag{9}$$

Recall that the nominal contact region is part of the mean plane. Therefore, $\eta$ is in fact equal to the asperity density. We compute the area of the nominal contact region formed after the occurrence of the large scale "pull-in" instability as

$$\Delta A_c = A_c^{h_{\min}} - A_c^0, \tag{10}$$

where $A_c^0$ and $A_c^{h_{\min}}$ are areas of the nominal contact region at the large-scale "pull-in" instability point (*i.e.*, $h = 0$, marked as state *ii* in Fig. 1 (b)) and at the maximum indentation depth (*i.e.*, $h = h_{\min}$, see Fig. 1 (b)), respectively. Our rough contact does not provide predictions for the nominal contact region. In many contact experiments the nominal contact region is measured as part of the experiment (*e.g.*, see Guduru and Bull[11]). In such cases, the total number of depth-dependent asperities can be estimated by using the measured nominal contact area values in conjunction with eqs. (9) and (10). In other situations, where such measurements are unavailable we believe that the best alternative is to estimate the nominal contact region using a classical, adhesive elastic contact theory. For example, in the next section we estimate the nominal contact region in the experiments of Kesari *et al.* using the JKR theory.

We estimated the energy loss for a single depth-dependent asperity, $\Delta E_{\text{md}}$, using the MD theory. That energy loss is not constant between the asperities, but varies between them depending on their curvature. Using eq. (7), we find that the variation of curvatures in the population of all asperities contained in any unit region of the mean plane to be

$$p_\kappa(t) = \sqrt{\frac{3}{4\pi}}\left(t^2 - 2 + 2e^{-\frac{t^2}{2}}\right)e^{-\frac{(8C_1^2 - C_2^2)t^2}{16C_1}}. \tag{11}$$

Recall that the nominal contact region belongs to the mean plane. Therefore, the PDF eq. (11) also applies to the population of all the asperities contained in any nominal contact region of unit area. Since the depth-dependent asperities are the total number of asperities contained in the nominal contact region formed after the occurrence of the large scale "pull-in" instability, the PDF eq. (11) also applies to the population of all depth-dependent asperities. Thus, the mean energy loss per depth-dependent asperity can be computed as

$$\langle \Delta E_{\text{md}} \rangle = \int_{-\infty}^{0} \Delta E_{\text{md}} p_\kappa(t)\,dt. \tag{12}$$

Writing $\Delta E_{\text{md}}$ in eq. (12) in terms of $\Delta \bar{E}_{\text{md}}$ using eq. (4a), and then approximating $\Delta \bar{E}_{\text{md}}$ with $\Delta \tilde{E}_{\text{md}}$ defined in eq. (6), we get

$$\langle \Delta E_{\text{md}} \rangle \approx 2E^* \int_{-\infty}^{0} \Delta \tilde{E}_{\text{md}}(\chi, \ell_0(t)) p_\kappa(t) R^3(t)\,dt, \tag{13}$$

where

$$\ell_0(t) = \frac{\pi w}{E^* R(t)}, \tag{14a}$$

$$R(t) = \left(\frac{1}{R_t} - \sqrt{\frac{m_4}{3}} t\right)^{-1}. \tag{14b}$$

Equation (14a) follows from noting that in eq. (1) the second argument, $\ell_0$, of the function $\Delta \tilde{E}_{\text{md}}$ depends on the effective mean curvature $1/R$, which is the sum of mean curvatures of the solids at their respective points of contact. We assume that at all contact points the tip's curvature equals $1/R_t$ and that the asperity's curvature at the contact point is the same as its curvatures $k_m$ at its apex. Equation (14b) follows these assumptions and the fact that $k_m = -\sqrt{m_4/3} t$.

Multiplying the mean energy loss per depth-dependent asperity with the total number of those asperities we get,

$$\Delta E_D = \eta \Delta A_c \langle \Delta E_{\text{md}} \rangle, \tag{15}$$

where $\eta$, $\Delta A_c$, and $\langle \Delta E_{\text{md}} \rangle$ are, respectively, given by eqs. (9), (10), and (13).

## Comparison with experiments

Equation (15) applies to arbitrary homogeneous and isotropic rough surfaces. In this section, we use eq. (15) to estimate $\Delta E_D$ in the glass bead–PDMS contact experiments reported by Kesari *et al.*[6] and compare the estimates with measured values. The experiments involved contact between a spherical glass bead and PDMS substrates. The geometry of the contacting solids in the experiments is shown in the insets of Fig. 2 (a). The radius of the glass bead (tip) was $R_t = 25$ $\mu$m and the PDMS substrates were 3–5 mm thick. In the experiments both the substrate and the tip are rough [see Figs. 7 (a–c)]. However, in our model, we assumed that only the substrate was rough. This makes the quantitative comparison of our model with experiments challenging. Nevertheless, we still attempt to compare our model's predictions with experiments by simply ignoring the tip's roughness and assuming the tip to be smooth. We believe that some knowledge can yet be gained about the utility of our model from such a comparison.

To estimate $\Delta E_D$ from eq. (15) we need to know the mean energy loss per asperity $\langle \Delta E_{\text{md}} \rangle$, the asperity density $\eta$, and the nominal contact area $\Delta A_c$ in the context of the experiments. We discuss the computation of each of these quantities in the following paragraphs.

The mean energy loss $\langle \Delta E_{\text{md}} \rangle$ can be computed from eqs. (13)–(14) on knowing the parameter $\chi$, the values of $\ell_0$, $p_\kappa$, and $R$ for a given $t$. To compute $\chi$ we need the values of physical parameters $E^*$, $w$, and $Z_0$. The material properties $E^*$ and $w$ were measured in the experiments to be 0.75 MPa and 26 mJ/m$^2$, respectively. However, we could not find a clear way to identify $Z_0$ in the experiments. The parameter $Z_0$ is a measure of the distance of the inter-body cohesive forces. This distance has been found in other experiments to range from 10 nm to 1 $\mu$m (see Table. 1). Therefore, we estimate $\Delta E_D$ for a number of different $Z_0$ values lying in that range.

The PDF $p_\kappa$ is completely defined by the spectral moments, $m_n$ ( $n = 0$, 2 and 4). We compute the spectral moments from eq. (8) after determining the isotropic, PSD function $C^{\text{iso}}$ in the experiments. The function $C^{\text{iso}}$ is a surface property and is determined by the surface's topography function $z$. As mentioned previously, we ignore the tip's roughness and only compute $C^{\text{iso}}$ for the substrate's surface. Kesari *et al.* reported that it was difficult to measure the PDMS substrates' nanometer scale surface topography because of the PDMS's low stiffness. As an alterative, they assumed that the salient features of a PDMS substrate's topography were well approximated by that of the Si mold on which it was cast. Following them, we too approximate the PDMS substrate's topography by that of the Si mold on which it was cast. Kesari *et al.* cast PDMS substrates on four different Si molds having different surface topographies. The RMS roughness $\sigma$ of those topographies ranged between 0.65 and 1.52 nm [see, *e.g.*, Fig. 2 (b)]. Since our's is a large surface roughness model, we only consider the experiments that correspond to the Si mold with the largest roughness, *i.e.*, the one corresponding to $\sigma = 1.52$ nm. Figure 7 (c) shows the surface topography of that Si mold. Using that topography data and the method presented in Ref.[36] we numerically computed the value of $C^{\text{iso}}$ in the experiments for a discrete set of wavenumber magnitudes. See Supplementary Material for details. The numerically computed $C^{iso}$ values are shown in Fig. 7 (d). As can be seen in the figure, the values of $C^{iso}$ are approximately constant at small wavenumber magnitudes, and fall off quickly at large wavenumber magnitudes. This behavior is similar to that of a *power-law* PSD function. To be specific, consider the PSD function

$$C^{iso}(q) = \frac{1}{C_0} \frac{e^{-q/q_0} L^2 \sigma^2}{(1 + L^2 q^2)^n}, \tag{16}$$

where

$$C_0 = 2\pi \int_0^\infty \frac{e^{-q/q_0} L^2 q}{(1 + L^2 q^2)^n} \, dq,$$

and $L$, $q_0$, and $n$ are parameters referred to as the correlation length, cut-off wavenumber, and the power-law index, respectively. The PSD function (16) is a modified version of the *k-correlation*, or *ABC* model, which has been shown to be applicable to a large variety of surface topographies[37,38]. We found that eq. (16) describes the numerically computed values of $C^{iso}$ in the experiments remarkably well for the parameter values $q_0 = 1.0 \times 10^8$ m$^{-1}$, $\sigma = 1.41$ nm, $L = 16.2$ nm, and $n = 3.28$ [Fig. 7 (d)]. The values for $\sigma$, $L$, and $n$ were obtained by minimizing a measure of the difference between the numerically computed values of $C^{iso}$ in the experiments and the values given by eq. (16). The value for $q_0$ was, however, chosen independently before performing the minimization. The value of 1.41 nm for $\sigma$ that we obtained through the minimization process is quite close to the value of 1.52 nm that Kesari *et al.* obtained from experimental measurements. This close match reinforces our interpretation that, for our chosen and fit parameter values, the $C^{iso}$ function given by eq. (16) approximates the experimental one well. Therefore, we use it in eq. (8) to estimate the spectral moments in the experiments. We get those estimates to be $m_0 \approx 2.31$ nm$^2$, $m_2 \approx 3.24 \times 10^{-5}$, and $m_4 \approx 4.02 \times 10^{-9}$ nm$^{-2}$. The $\alpha$ value corresponding to these estimates is 8.85. Using these estimated values we can compute an approximate value for $p_\kappa$ in the experiments for any given $t \in [0, \infty)$.

**Table 1.** Estimates for the range of $Z_0$ from literature

| Materials | Geometry | Range |
|---|---|---|
| Silica–Silica[33] | Sphere (of radius 3.8 $\mu$m)–Plate | 10 nm |
| Polystyrene–Glass[34] | Sphere (of radius 6 $\mu$m)–Plate | 20–100 nm |
| Glass–Glass[35] | Plate–Plate | 0.68–1.2 $\mu$m |

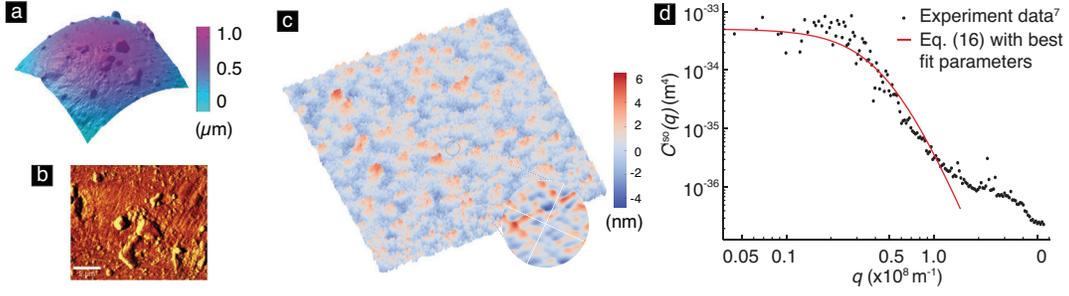

**Figure 7.** (a) The curved shape of the glass bead and (b) its surface topography after subtracting the curvature effect. (c) The surface topography of the Si mold scanned over a area of 2 $\mu$m × 2 $\mu$m with a total of 256 points in each direction. The measured RMS roughness $\sigma$ is 1.52 nm. (d) The power spectrum of the Si mold's surface topography, and the corresponding fitting to the power-law PSD function eq. (16) with best fitting parameters $\sigma = 1.41$ nm, $L = 16.2$ nm and $n = 3.28$.

Similarly, we can compute approximate values for $R$, and consequently for $\ell_0$, in the experiments for any given $t \in [0, \infty)$. Using those approximate values in eq. (13), we find that $\langle \Delta E_{md} \rangle$ ranges from $2.5 \times 10^{-4}$ pJ to 0.31 pJ as $Z_0$ varies from 1 $\mu$m ($\chi = 0.0347$) to 10 nm ($\chi = 3.47$). Using the estimated values for the spectral moments and eq. (9) we found the experimental value of $\eta \approx 3.8 \, \mu$m$^{-2}$.

The final quantity needed to estimate $\Delta E_D$ in the experiments is the nominal contact area $\Delta A_c$. Unfortunately, Kesari *et al.* do not report measurements of the nominal contact area in their experiments. As the next best alternative, it would have been ideal if our model gave predictions for $\Delta A_c$, which it does not. For a lack of a better alternative, we use the JKR theory to estimate $\Delta A_c$ in the experiments. The JKR theory is the most widely used model for adhesive elastic contact, which only applies to contact between smooth surfaces. Kesari and Lew[16] presented a generalization of the JKR theory that applies to contact between rough surfaces and gives a prediction for the nominal contact area. However, as we discussed in Introduction, that model only applies in the regime of small surface roughness. Employing the JKR theory we find that $\Delta A_c$ in the experiments is approximately equal to $4R_t|h_{min}|$. See Supplementary Material for details.

Figure 8 (a) shows the $\Delta E_D$ values that we obtained using our model's estimates for $\langle \Delta E_{md} \rangle$, $\eta$, and $\Delta A_c$. It also shows the experimentally measured $\Delta E_D$ values reported by Kesari *et al.* Recall that $Z_0$ is an unkown in the experiments of Kesari *et al.*, so we use it as a free parameter while generating the predictions from our model. We find that our model's predictions match the experimental measurements quite well when $Z_0$ is around 520 nm [see Fig. 8 (a)]. As can be seen from Table 1, this value for $Z_0$ is well within the range of the values measured for $Z_0$ in other experiments.

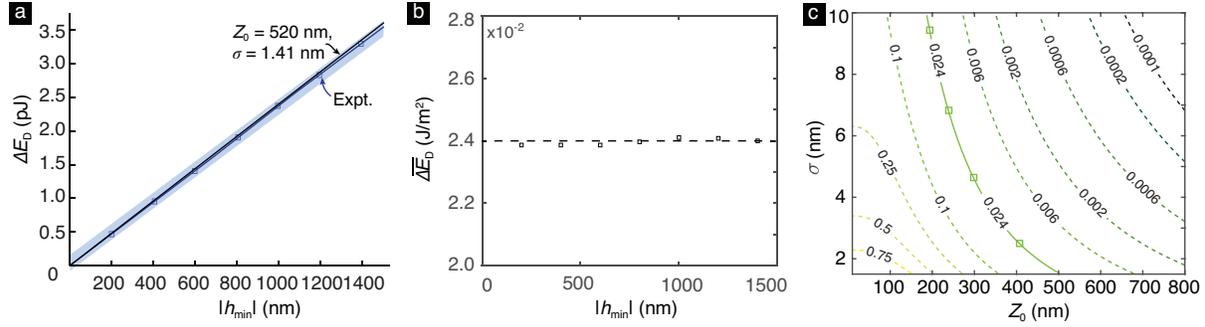

**Figure 8.** (a) The comparison of the depth-dependent energy loss $\Delta E_D$ measured in the experiment[6] with the estimations based on the model according to eq. (15) for $Z_0 = 520$ nm and $\sigma = 1.41$ nm. The shaded region indicates the standard deviation of the measurements. (b) The plot of depth-dependent energy loss per unit nominal contact area $\overline{\Delta E}_D$ with $|h_{\min}|$. (c) The contour plot of the depth-dependent energy loss per unit area $\overline{\Delta E}_D$ as a function of $Z_0$ and $\sigma$. The experimental value of $\overline{\Delta E}_D$ is $2.4 \times 10^{-2}$ J/m$^2$ and is shown as square symbols.

Consider the quantity

$$\overline{\Delta E}_D = \frac{\Delta E_D}{\Delta A_c} \approx \frac{\Delta E_D}{4 R_t |h_{\min}|}, \qquad (17)$$

which is the depth-dependent energy loss per unit nominal contact area. This quantity is a constant in our model. Since, in it $\Delta E_D$ depends linearly on $|h_{\min}|$ on account of $\Delta E_D$ depending linearly on $\Delta A_c$, and $\Delta A_c$ depending linearly on $|h_{\min}|$. Figure 8 (b) shows the values of $\overline{\Delta E}_D$ in the experiments at different $|h_{\min}|$. We computed these values using the data shown in Fig. 8 (a). As can be seen, $\overline{\Delta E}_D$ is essentially a constant with respect to $|h_{\min}|$ in the experiments. Thus, our model's prediction that $\Delta E_D$ varies linearly with $|h_{\min}|$ is in good agreement with experimental measurements.

## Effect of $\sigma$ and $Z_0$ on $\Delta E_D$

We found that in our model the inter-body interaction length-scale and the surface roughness have significant effect on $\Delta E_D$. We studied the effect of $\sigma$ by scaling the $z$ measurements of Kesari *et al.* [Fig. 7 (b)] by different factors and repeating all the calculations that we described in the previous section. (Scaling the $z$ measurements by a factor, say $k$, changes $\sigma$ by a factor of $k^2$.) The results from those calculations for $\overline{\Delta E}_D$ are shown in Fig. 8 (c). As can be seen, $\overline{\Delta E}_D$, and hence $\Delta E_D$, decreases and tends to zero as $\sigma$ increases. This behavior is in agreement with the trend of $\Delta E$ decreasing with roughness at large surface roughnesses reported by Kesari *et al.* and others, as discussed in Introduction.

Also seen in Fig. 8 (c) is the trend that $\overline{\Delta E}_D$, and hence $\Delta E_D$, decreases as the adhesive interaction length-scale $Z_0$ increases. We are unaware of any experimental data that can be used to check the validity of this theoretical prediction of our model. However, this trend is consistent with the numerical results reported by Song *et al.*[24], in which the strength of adhesion decreases as the adhesive interaction length-scale increases.

## Concluding remarks

We generated predictions from our model in the context of the experiments reported by Kesari *et al.* In general, however, it is challenging to determine *a priori* whether or not it is reasonable to apply our model to a particular contact scenario. The reason for this is as follows. We assumed in our model that the contact region is multiply connected and that there is no interaction between neighboring asperities. These are reasonable assumptions only if the size of the contact region formed at each asperity is much smaller than the separation between neighboring asperities. However, we are not aware of any general criteria/models that can be used to gather information in this regard without actually solving for the complex stress and displacement fields at the contact interface. Therefore, a general theory of the type developed by Johnson[39] that yields information about the topology of the contact region would form a valuable supplement to our model.

We conclude by noting that our model bears some similarities with the models presented in Ref[40,41]. In particular, following Fuller and Tabor's[23] approach, Wei *et al.*[40] investigated the effect of roughness on adhesion hysteresis. However, there are significant differences between their and our models. For example, Wei *et al.* assumed the asperities' radii of curvatures to be a constant, whereas in our model the asperities have different radii of curvatures depending on their heights. In Wei *et al.*'s model the asperity level contact is modeled using the JKR theory, whereas we model that interaction using the MD theory.

Finally, Wei *et al.*'s model does not capture the depth-dependent nature of the hysteretic energy loss. Our model provides a semi-analytical formula to estimate the depth-dependent energy loss.

## Acknowledgements

This work was supported by the National Science Foundation through the Mechanics of Materials and Structures Program award number 1562656. Weilin Deng is partially supported by a graduate fellowship from the China Scholarship Council.

## Author contributions

H.K. designed the research. W.D. and H.K. developed the mechanics model, analyzed the results, and wrote the manuscript.

## Additional information

**Supplementary information** of this manuscript can be found in the Supplementary Material.
**Competing Interests**: The authors declare that they have no competing interests.